# Optimum gain for plasmonic distributed feedback lasers

A. A. Zyablovsky,[1,2] I. A. Nechepurenko,[1,2] E. S. Andrianov,[1,2] A. V. Dorofeenko,[1,2,3], A. A. Pukhov,[1,2,3] A. P. Vinogradov,[1,2,3] and A. A. Lisyansky[4,5]

[1]Dukhov Research Institute for Automatics, 22 Sushchevskaya, Moscow 127055, Russia

[2]Moscow Institute of Physics and Technology, 9 Institutskiy per., Dolgoprudniy 141700, Moscow Reg., Russia

[3]Institute for Theoretical and Applied Electromagnetics RAS, 13 Izhorskaya, Moscow 125412, Russia

[4]Department of Physics, Queens College of the City University of New York, Queens, NY 11367, USA

[5]The Graduate Center of the City University of New York, New York, New York 10016, USA

Although nanolasers typically have low *Q*-factors and high lasing thresholds, they have been successfully implemented with various gain media. Intuitively, it seems that an increase in the gain coefficient would improve the characteristics of nanolasers. For a plasmonic nanolaser, in particular, a distributed-feed-back (DFB) laser, we propose a self-consistent model that takes into account both spontaneous emission and the multimode character of laser generation to show that for a given pumping strength, the gain coefficient has an optimal value at which the radiation intensity is at a maximum and the radiation linewidth is at a minimum.

## 1. INTRODUCTION

Laser physics has been extensively developed since the discovery of the laser more than half a century ago,[1-3] and now the principles of its operation have been well established. A laser consists of a resonator and a gain medium, in which a population inversion is created by external pumping.[4] Spontaneous emission with population inversion results in an increase of the number of photons inside the resonator. The resonator provides a positive feedback stimulating the gain medium to radiate into the resonator modes. The gain medium, as an amplifier, together with the feedback makes a coherent light generator.

An amplifying medium is usually described by the gain coefficient *G*, which does not depend on the pump rate. This coefficient, $G = n_G \sigma_G$, is the product of the concentration of atoms of the active medium, $n_G$, and the cross section of their interaction with the electromagnetic wave, $\sigma_G$;[5-7] $\exp(G)$ is the factor by which the amplitude of the plane wave is increased after its propagation through a unit length of the amplifying medium. In traditional lasers the gain coefficient varies in the range of 0.1–1 cm$^{-1}$.[8]

Lasing begins when the pump power exceeds a threshold at which the energy supplied compensates the loss. Energy is lost in the resonator walls and to radiation of light including laser output. There are several ways of increasing the laser power. One can either increase the pump power or decrease losses. Decreasing losses results in increasing the $Q$-factor of the resonator and thereby decreasing the lasing threshold. The threshold decrease may also be achieved by using new materials with enhanced gain. This often leads to emergence of lasers with a new design. In traditional macroscopic lasers, however, lasing can be achieved at fairly low values of the gain coefficient and a reasonable pump power. Therefore, there is no need for special efforts to increase the gain coefficient.

In the last decade, a new class of lasers, nanolasers, has become a subject of ever-growing interest because of the prospect of creating a subwavelength source of coherent radiation.[9-19] Owing to its small size, the nanolaser allows for ultrafast field dynamics.[13, 15] This leads to the possibility of ultrafast transceivers that are important for developing optoelectronic technologies that require sources of coherent radiation that are capable of rapid transformations of electrical to optical signals. However, the side effect of the subwavelength size of plasmonic nanolasers is a wide angular distribution of the emitted light. To narrow the directional distribution, at least one of the laser dimensions should be of the wavelength size. For this purpose, one can use conically shaped nanoantennas,[20] Yagi-Uda antennas,[21, 22] or periodic chains[23] and lattices of nanoparticles. The latter forms a plasmonic DFB nanolaser,[17-19, 24-34] which is the subject of our paper.

In nano DFB lasers, the role of the resonator is played by the periodic plasmonic structure. These can be metallic films perforated by holes or slits[18, 19, 24, 26, 28, 32] and one- or two-dimensional arrays of plasmonic nanoparticles.[25, 27, 30] The eigenmodes of plasmonic DFB lasers are hybrid plasmon-polariton Bloch modes.[18, 25, 27] These modes can be generated with either continous[18, 24, 28, 32] or pulsed[25-27, 29-31, 33, 34] pumping.

Even though substantial progress has been made in creating plasmonic DFB lasers,[18, 24-34] a number of problems remain unsolved. The main problem with a DFB laser is that a plasmonic system has high losses and, thereby, a low $Q$-factor and a high value of the first lasing threshold. Consequently, in these systems, the contribution of spontaneous emission is high. Therefore, plasmonic DFB lasers may be overheated at levels of pump power required to achieve coherent lasing.[35-37] High-gain materials have been used to address this challenge.[18, 28] These are either semiconductor quantum wells[18, 24, 28, 32] or organic dyes[25-27, 29-31, 33, 34] for which gain coefficients are in the range of 100-1,000 cm$^{-1}$. These materials, however, are either expensive or require cryogenics for laser operations. Therefore, the search for low-loss plasmonic materials and active materials with high gain continues.[38-41]

In typical high-$Q$ lasers, multimode excitations may result in various effects. These include mode competition, which may suppress lasing from all modes except from the strongest one, mode locking, and mode beating, which are well studied in the literature.[12, 42, 43] In plasmonic DFB lasers, mode competition is suppressed by low *Q*-factors. Therefore, these systems are essentially multimode. In this paper, we consider a synergetic cooperation of spontaneous emission and a multimode regime.

We show that an increase in the gain coefficient would not necessarily improve nanolaser output. There is an optimal value of the gain that results from the nonlinear interaction of modes via the gain medium. For this value, a maximum intensity of radiation for a given pump power is realized. If the gain exceeds the optimal value, the radiation intensity decreases and the laser linewidth increases. We study the effect of spontaneous emission on multimode generation in a DFB nanolaser and do not consider effects of nonuniform distributions of a gain medium[42, 44] and a pump power in the resonator,[45-49] as well as the non-monotonic dependence of the laser generation threshold on parameters of the resonator.[50-52]

## 2. THE SYSTEM

In experiments, to create an inverted population of an amplifying medium, an optical pulse pumping is usually used.[25-27] The duration of the pulse varies from 40 fs to 200 ns.[25, 26] A pulse must not overheat and destroy an active medium, which limits the pulse duration. A characteristic time of the longitudinal relaxation of active medium atoms is less than 10 ns. If a pump pulse has a duration of 100 ns or more, then in a laser stationary oscillations are established. Such a pulse, therefore, can be approximately considered a CW regime. This is the regime that we consider below.

To demonstrate the concept, we consider a DFB laser in which plasmonic nanoparticles are periodically positioned in a layer of a gain medium (see Fig. 1). A similar system was realized experimentally in Ref. 27.

Eigenmodes of such a laser are the Bloch modes, whose wavenumbers $k_B$ lie in the plane parallel to the gain layer and satisfy the phase condition of the laser generation[50, 55]

$$L \operatorname{Re} k_B = \pi n + \arg r, \tag{1}$$

where *L* is the length of the laser side, *n* is an integer, and *r* is the coefficient of reflection from the laser boundary. This coefficient is determined by both the laser structure and properties of the surrounding medium. We assume that $\arg r = 0$. For calculation of eigenmodes of the plasmonic DFB laser, we apply Bloch boundary conditions to the elementary cell of the periodic lattice. By going through the Bloch wavenumbers satisfying Eq. (1), we find the corresponding complex eigenfrequencies, $\omega + i\gamma$ shown in Fig. 2.

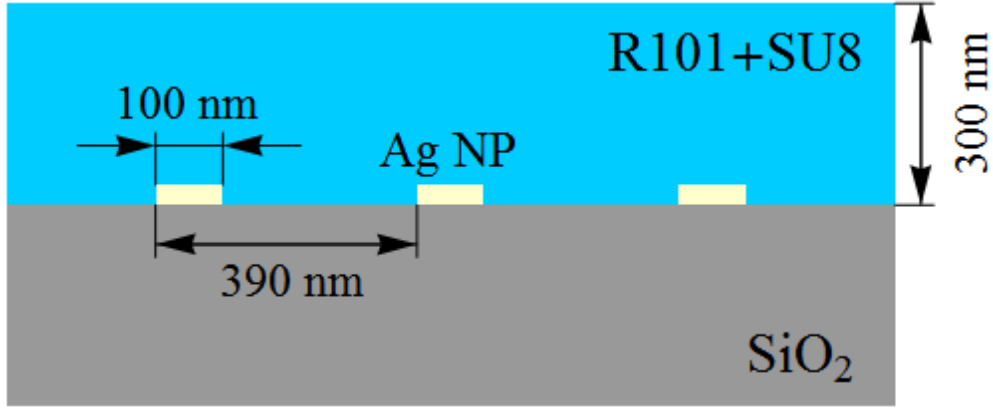

Fig. 1. Schematics of the DFB laser. This system consists of a 300-nm layer of a mixture of polymer SU8 and dye R101 that is applied on a quartz substrate. Inside the layer, cylindrical silver nanoparticles are placed. The diameter and the height of a nanoparticle are 100 nm and 30 nm, respectively. The nanoparticles are positioned in periodic square-lattice sites with the period of $l = 390$ nm, the size of the system is $L = 10l$. The transition frequency of dye molecules is $\omega_\sigma = 3.195 \cdot 10^{15}$ rad/s and their longitudinal and transverse relaxations rates are $\gamma_d = (4.3 \text{ ns})^{-1}$ and $\gamma_\sigma = (4.6 \text{ fs})^{-1}$, respectively.[53, 54]

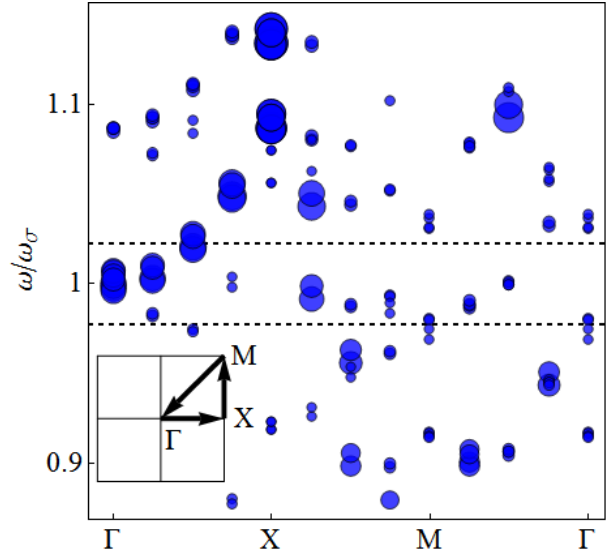

Fig. 2. Eigenmodes of a plasmonic DFB laser at the coordinates ΓXMΓ. The size of a dot reflects the value of $Q$-factor, $\omega/(2\gamma_i)$, of the respective mode. Horizontal dashed lines show the amplification linewidth of dye molecules. The inset shows the irreducible Brillouin zone with the ΓXMΓ path.

Note that the radiation from the surface of the DFB laser arises if the spatial spectrum of the generated Bloch wave contains wavenumbers which tangential components are smaller than the wavenumber of the surrounding space.[18, 24, 27] Below we consider the case when the radiation is in the direction perpendicular to this plane. Despite the small thickness of the active layer the width of the radiation pattern of the emitter may be narrow because it depends on the size of the radiation aperture which, in our case, is the transverse size of the laser.

## 3. MATHEMATICAL MODEL

A consistent study of plasmonic DFB lasers should take into account the multimode character of laser generation and spontaneous radiation of "atoms" of the gain medium. There are two equivalent approaches to the description of spontaneous radiation in multimode lasers. The first one is based on the solution of the operator version of the Maxwell-Bloch equations – the Heisenberg-Langevin equations. These equations include noise terms, thanks to which they take into account spontaneous emission processes.[56, 57] However, operators describing noise in the Heisenberg-Langevin equations, are non-commutative. Therefore, working with them is difficult. For a large number of atoms, operator equations can be replaced by equations for $c$-numbers to obtain Maxwell-Bloch equations.[58] Such an approach is commonly used for the consideration of laser generation[59-62] including studies of multimode lasers.[63]

In our paper, we use another approach, which is based on equations for expectation values for operators of the number of photons, the population inversion of atoms of the gain medium, and operators of coupling between these atoms and the electromagnetic field.[16, 64-67] These equations can be obtained by using the Lindblad form[68, 69] of the master equation for the density matrix (see Appendix A). Such a system of equations does not contain a noise term because the master equation for the density matrix is deterministic. The main problem in such an approach is an infinite number of equations. However, in many problems it is possible to uncouple some operator correlators to obtain a finite system. For a laser system, a common approach is uncoupling of the correlator of the number of photon, $\hat{a}^+\hat{a}$, and the population inversion, $\hat{D}$, namely, $\langle \hat{a}^+\hat{a}\hat{D} \rangle = \langle \hat{a}^+\hat{a} \rangle \langle \hat{D} \rangle$, which results in a finite number of coupled nonlinear equations that take into account spontaneous emission (see Ref. 64). In the limit of a large number of atoms, the difference between $\langle \hat{a}^+\hat{a}\hat{D} \rangle$ and $\langle \hat{a}^+\hat{a} \rangle \langle \hat{D} \rangle$ falls off as the inverse of the number of atoms,[68] which justifies the applicability of such an approximation for DFB lasers. This approach is suitable for both numerical simulations and analytical evaluations, and it is used in our paper.

To derive the laser equations that take into account both spontaneous radiation and multimode character of the laser generation, we use the Jaynes–Cummings Hamiltonian:[57, 68]

$$\hat{H} = \sum_i \hbar\omega_i \hat{a}_i^+ \hat{a}_i + \sum_m \hbar\omega_\sigma \hat{\sigma}_m^+ \hat{\sigma}_m + \sum_{i,m} \left( \hbar\Omega_{im} \hat{a}_i^+ \hat{\sigma}_m + \hbar\Omega_{im}^* \hat{a}_i \hat{\sigma}_m^+ \right), \qquad (2)$$

where $\hat{a}_i^+$ and $\hat{a}_i$ are creation and annihilation operators of a quantum in the $i$-th resonator mode, $\omega_i$ is the frequency of this mode, $\hat{\sigma}_m^+$ and $\hat{\sigma}_m$ are raising and lowering operators for the transition of the *m*-th two-level atom, $\omega_\sigma$ is the frequency of this transition, and $\Omega_{im}$ is the coupling constant between the field in the *i*-th mode and the *m*-th atom.

Using Hamiltonian (2) and the master equation, Eq. (A1), we obtain equations for the expectation values of the operator of the number of photons in the *i*-th mode, $\hat{n}_i = \hat{a}_i^+ \hat{a}_i$, the operator of the population inversion of the *m*-atom, $\hat{D}_m = \hat{\sigma}_m^+ \hat{\sigma}_m - \hat{\sigma}_m \hat{\sigma}_m^+$, the operator of the energy flux from the *m*-th atom to the *i*-th mode,[16, 64] $\hat{I}_{im} = -i\left(\hbar\Omega_{im}\hat{a}_i^+\hat{\sigma}_m - \hbar\Omega_{im}^*\hat{a}_i\hat{\sigma}_m^+\right)$, and the interaction operator between the *m*-th atom and the field in the *i*-th mode, $\hat{V}_{im} = \hbar\Omega_{im}\hat{a}_i^+\hat{\sigma}_m + \hbar\Omega_{im}^*\hat{a}_i\hat{\sigma}_m^+$ [Eqs. (A6)-(A9)]:

$$\frac{dn_i}{dt} = -\gamma_i n_i + \sum_m I_{im}, \tag{3}$$

$$\frac{dD_m}{dt} = -\gamma_d \left(D_m - D_m^0\right) - 2\sum_i I_{im}, \tag{4}$$

$$\frac{dI_{im}}{dt} = -\left(\gamma_\sigma + \gamma_i/2\right) I_{im} + \left(\omega_i - \omega_\sigma\right) V_{im} + |\Omega_{im}|^2 \left(2n_i D_m + D_m + 1\right), \tag{5}$$

$$\frac{dV_{im}}{dt} = -\left(\gamma_\sigma + \gamma_i/2\right) V_{im} + \left(\omega_\sigma - \omega_i\right) I_{im}, \tag{6}$$

where $\gamma_i$ is the relaxation rate of the photons in the *i*-th resonator mode and $\gamma_d$ and $\gamma_\sigma$ are the rates of the longitudinal and transverse relaxations in an atom. The term $\gamma_d D_m^0$ describes processes of incoherent pumping of the two-level active medium.[70]

The system of equations (3)-(6) describes interactions between each resonator mode and each atom of the amplifying medium. If we assume that the constants of interactions of the *i*-mode with each atom are the same, $\Omega_{im} = \Omega_i$, then, in Eqs. (3)-(6), we can sum up over all atoms of the amplifying medium. With this assumption, the Maxwell-Bloch equations predict a single-mode regime for any gain.[64] As a result, we obtain:

$$\frac{dn_i}{dt} = -\gamma_i n_i + I_i, \tag{7}$$

$$\frac{d\bar{D}}{dt} = -\gamma_d \left(\bar{D} - \bar{D}_0\right) - \frac{2}{N}\sum_i I_i, \tag{8}$$

$$\frac{dI_i}{dt} = -\left(\gamma_\sigma + \gamma_i/2\right) I_i + \left(\omega_i - \omega_\sigma\right) V_i + |\Omega_i|^2 \left(2n_i \bar{D} + \bar{D} + 1\right), \tag{9}$$

$$\frac{dV_i}{dt} = -\left(\gamma_\sigma + \gamma_i/2\right) V_i + \left(\omega_\sigma - \omega_i\right) I_i, \tag{10}$$

where $\bar{D}$ is the average population inversion of atoms of the amplifying medium, $I_i$ is the total energy flux from all atoms of the amplifying medium into the *i*-th resonator mode, $V_i$ is the energy of the interaction between the *i*-th resonator mode and all atoms of the amplifying medium, and $N$ is the number of atoms. Note, that energy loss in the resonator is due to both losses in the material and radiation. Therefore, for each mode, the relaxation rate $\gamma_i$ can be represented as a sum of rates due to non-radiation, $\gamma_{NRi}$, and radiation losses, $\gamma_{Ri}$.

The Jaynes–Cummings Hamiltonian that was used for deriving Eqs. (7)-(10) describes the interaction of the field with two-level atoms. In reality, amplifying media have complicated multi-level structures. Often, a multi-level system can be approximately considered as a three- or four-level system which, in turn, can be reduced to an effective two-level system.[70] Below we assume that the concentration of atoms of the active medium is $n_G = 5 \cdot 10^{17}$ cm$^{-3}$, which allows us to neglect nonradiation interactions among the atoms of an amplifying medium (e.g., the Förster transitions).

Note, that in high *Q*-factor lasers, $\gamma_\sigma \gg \gamma_i$. Therefore, variables $I_i$ and $V_i$ can be adiabatically excluded from Eqs. (7)-(10). As a result, these equations become identical to the rate equations for the number of photons and the population inversion. In low-quality resonators, such as plasmonic resonators that are considered here, the condition $\gamma_\sigma \gg \gamma_i$ is not fulfilled for some modes. In this case, Eqs. (7)-(10) cannot be reduced to the rate equations, and one has to solve the system (7)-(10).

The coupling constant between the *i*-th resonator mode and the amplifying medium, $|\Omega_i|^2$, is proportional to $G$ [see Appendix B, Eq. (B12)]:

$$|\Omega_i|^2 \sim \eta_i G, \qquad (11)$$

where $\eta_i = W_E / W_T$, $W_E$ is the energy of the electric field of the *i*-th eigenmode in the volume of the amplifying medium, $W_T$ is the total energy of this mode:

$$\eta_i = \frac{W_E}{W_T} = \frac{\frac{1}{8\pi} \int_{V_G} \left[ \partial \mathrm{Re}(\varepsilon\omega)/\partial\omega \right]_{\omega=\omega_i} |\mathbf{E}_i|^2 d^3\mathbf{r}}{\frac{1}{8\pi} \int_V \left( \left[ \partial \mathrm{Re}(\varepsilon\omega)/\partial\omega \right]_{\omega=\omega_i} |\mathbf{E}_i|^2 + |\mathbf{H}_i|^2 \right) d^3\mathbf{r}}, \qquad (12)$$

where $\varepsilon(\mathbf{r})$ is the coordinate dependent dielectric permittivity of the DFB laser in which all media are considered as nonmagnetic. The dispersion of the dielectric permittivity must be taken into account for the correct calculation of the field in the structures with negative permittivities such as plasmonic structures.

As shown in Appendix B, the gain coefficient can be expressed as:

$$G = n_G \sigma_G = n_G \frac{\omega}{c} \frac{4\pi |\mathbf{d}|^2}{\hbar \gamma_\sigma \sqrt{\operatorname{Re}\varepsilon_G}}, \tag{13}$$

where $\mathbf{d}$ is the dipole moment of an atom, $\varepsilon_G$ is the dielectric permittivity of the active medium, and $c$ is the speed of light in vacuum.

## 4. COMPARISON OF THE DEVELOPED MODEL WITH THE SEMICLASSICAL MODELS OF LASING

By solving Eqs. (7)-(10) numerically we show that the number of photons in system modes is always non-zero. This number increases monotonically with an increase in the pump rate as shown in Fig. 3. In the stationary state, the number of photons in the $i$-th eigenmode is:

$$n_i(G) = \frac{1}{2} \frac{\alpha_i \eta_i G}{\gamma_i} \frac{\bar{D}(G)+1}{1-\alpha_i \eta_i G \bar{D}(G)/\gamma_i}, \tag{14}$$

the population inversion is

$$\bar{D} = \frac{\gamma_d \bar{D}_0 - \frac{1}{N} \sum_i \alpha_i \eta_i G}{\gamma_d + \frac{1}{N} \sum_i \alpha_i \eta_i G (2n_i + 1)}, \tag{15}$$

where

$$\alpha_i = \frac{c\gamma_\sigma^2}{\left(\gamma_\sigma^2 + (\omega_\sigma - \omega_i)^2\right)} \frac{\operatorname{Re}\sqrt{\varepsilon_G}}{2\pi \left(\partial \operatorname{Re}(\varepsilon_G \omega)/\partial \omega\right)}. \tag{16}$$

In Eqs. (9) and (14), spontaneous transitions are described by the term $(\bar{D}+1)$.

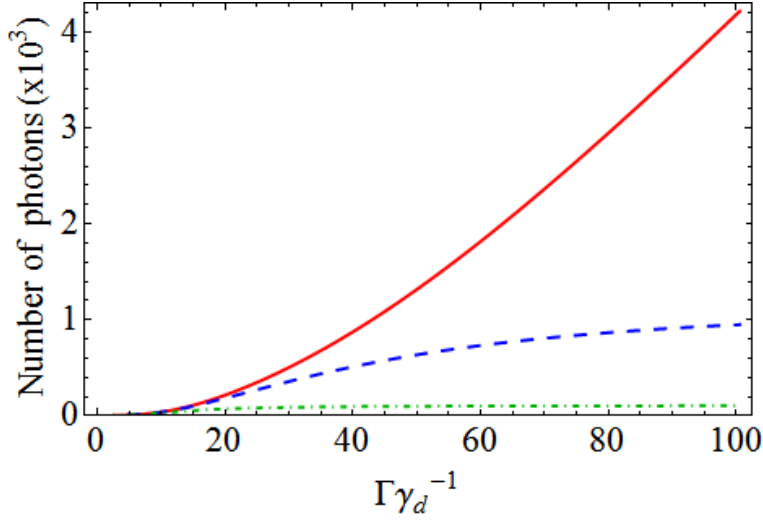

Fig. 3. The number of photons in different modes as a function of the pump rate of the active medium. The solid red line shows the number of photons in the mode with the lowest generation threshold, dashed blue and dash-dotted green lines show the numbers of photons in modes with the second and the third lowest generation thresholds, respectively. The gain coefficient is $G = 525$ cm$^{-1}$. In this and all other figures, the concentration of atoms of the active medium is $n_G = 5 \cdot 10^{17}$ cm$^{-3}$.

While the number of photons is small, we can assume that $\bar{D} \approx \bar{D}_0$. Using this approximation we can estimate the generation threshold as a function of $G$. The number of photons in the $i$-th mode starts rising sharply when the denominator $1 - \alpha_i \eta_i G \bar{D}_0 / \gamma_i$ in Eq. (14) approaches zero. One can, therefore, define the lasing threshold as

$$\bar{D}_{0i}^{th} = \gamma_i / \alpha_i \eta_i G. \tag{17}$$

This value coincides with the first generation threshold obtained from both the Maxwell-Bloch equations[57] and the rate equations when spontaneous radiation is not taken into account.[70]

However, away from this threshold, the results of our theory are different. Note that in the semiclassical Maxwell-Bloch theory due to the mode competition, photons exist in one mode only.[64] Moreover, the laser behavior does not change qualitatively when gain changes. This theory predicts that if $D_0 > \bar{D}_0^{th}$, then $\bar{D}(D_0) = \bar{D}_0^{th}$. The difference $\Delta_{MB}\bar{D} = \bar{D}_0 - \bar{D}(D_0) = D_0 - \bar{D}_0^{th}$ arises due to depletion of the inverse population $\bar{D}$ by stimulated emission. Our theory, Eqs. (14)–(16), takes into account spontaneous emission. For $\bar{D}_0 > \bar{D}_0^{th}$, the theory yields $\bar{D}(\bar{D}_0) < \bar{D}_0^{th}$ and $\Delta \bar{D} > \Delta_{MB}\bar{D}$. Additional depletion is caused by spontaneous emission and depends on gain. Moreover, spontaneous emission goes into all modes. Consequently, owing to spontaneous emission, there are photons in all modes, the stimulated emission occurs into all modes as well.

Below the generation threshold, the Maxwell-Bloch equations predict the absence of photons in the modes. According to Eqs. (7)-(10) for $\alpha_i \eta_i G \bar{D} / \gamma_i \ll 1$, the number of photons in each mode is proportional to $\alpha_i$ and is approximately the same for each mode (see Fig. 3). In this case, the DFB laser works in the multimode regime.

With an increase in the pump rate, the fastest growth of the number of photons is in the mode with the largest $\alpha_i$, i.e. in the mode with the lowest generation threshold. For $\alpha_i \eta_i G \bar{D} / \gamma_i > 1$, the number of photons in this mode is much greater than in all other modes. This is similar to the single-mode regime except for the dependence on $G$.

## 5. NON-MONOTONIC BEHAVIOR OF THE PHOTON NUMBER

In a real system, the characteristic gain parameter is $GD$. The population inversion $D$ depends on time and the pump rate. Assuming that the pump rate $\bar{D}_0$ is fixed, we analyze the dependence of numbers of photons in resonator modes on the gain coefficient $G$ of the active medium. From Eqs. (7)-(10), one can find that when the gain coefficient exceeds a threshold value $G_{th}$, the number of photons in any mode starts sharply increasing (Fig. 4). $G_{th}$ is defined by the condition (14):

$$G_{th} = \gamma_i / \alpha_i \eta_i \bar{D}_0. \tag{18}$$

Near the generation threshold, the population inversion of the active medium is approximately $\bar{D} \approx \bar{D}_0$, the number of photons is small and is approximately the same for all modes. Although, in this regime, the photon number in all mode increases with nearly the same rates with an increase in the gain coefficient (see inset in Fig. 4), in the mode with the lowest generation threshold, the rate increase is the greatest. The number of photons in the $i$-mode depends inversely on the value $\left( \bar{D}_{0i}^{th} - \bar{D} \right)$:

$$n_i = \frac{\bar{D}+1}{2\left( \bar{D}_{0i}^{th} - \bar{D} \right)}, \tag{19}$$

[see Eq. (14)]. Obviously, in the mode with the lowest generation threshold, $\bar{D}_{0i}^{th}$, $n_i$ has the maximum value.

An increase in gain $G$ results in a decrease of $\bar{D}_{0i}^{th}$ [see Eq.(17)] that in turn leads to an increase in the number of photons $n_i$, [see Eq. (19)]. At the same time, due to stimulated emission, the increase in the number of photons results in additional depletion of the population inversion of the active medium $\bar{D}$ leading to a slower photon rate increase above the threshold, $G_{th}$, in any mode. Because of the mode competition, a dramatic rate decrease occurs in any mode but the mode with the lowest generation threshold (see inset in Fig. 4).

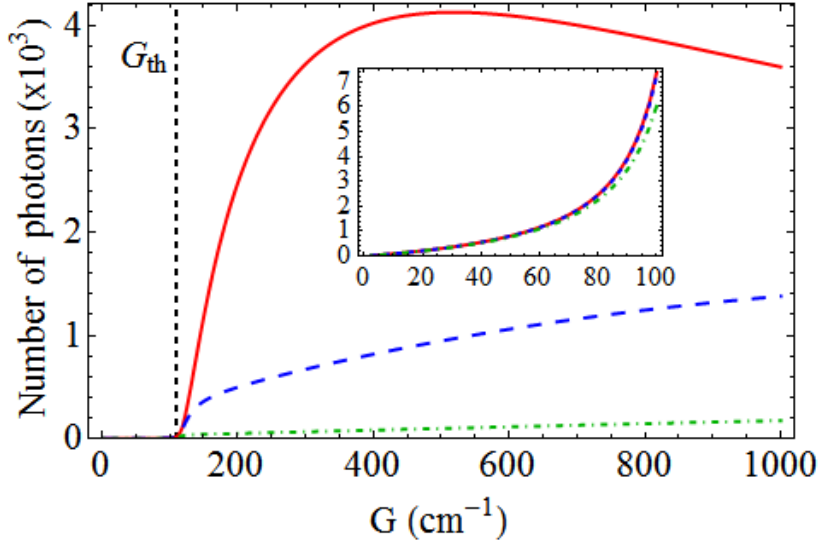

Fig. 4. The number of photons in different modes as a function of the gain coefficient of the active medium. The solid red line shows the number of photons in the mode with the lowest generation threshold, blue dashed and green dash-dotted lines show the numbers of photons in modes with the second and the third lowest generation thresholds, respectively. The inset shows the number of photons near the threshold gain coefficient, $G_{th}$, Eq. (18), which is shown by the vertical dotted line.

Our computer simulation based on the solution of Eqs. (7)-(10) shows that a further increase in the gain coefficient $G$ results in suppression of the lasing rate even in the mode with the lowest generation threshold. At a certain value of the gain coefficient, $G_{opt}$, the number of photons in this mode has a maximum (see Fig. 4).

In the stationary generation regime, the total rate of stimulated and spontaneous transitions into all modes cannot be greater than the pump rate. However until $\bar{D} > 0$, an increase in the gain coefficient $G$ results in an unlimited increase in the rate of spontaneous emission $\left( \Gamma_{sp}^i \sim |\Omega_i|^2 (\bar{D}+1) \sim \eta_i G(\bar{D}+1) \right)$ [see Eqs.(9) and (11)]. To stop the increase in the number of spontaneously emitted photons, the depletion of the population inversion should be sufficiently high to yield $\bar{D} = -1$. Note that, for $\bar{D} < 0$, the photon generation does not occur because stimulated transitions lead to photon absorption.

In a single-mode laser, for $\bar{D} < 0$, the lasing stops but the total number of photons does not decrease due to the growing number of spontaneously emitted photons. In a multi-mode laser, in the mode with the lowest generation threshold, the generation breakdown causes a decrease in both the number of coherent photons and the total photon number, because of the increase in spontaneous emission into the other modes.

Let us consider the change in the rate of spontaneous and stimulated transitions with a variation of the gain coefficient $G$. The rate of induced transitions is proportional to the number of photons in the mode,

$$\Gamma_{st}^{i} = 2\gamma_{\sigma}^{-1}|\Omega_{i}|^{2} n_{i}\bar{D} \sim 2\eta_{i} G n_{i}\bar{D}. \tag{20}$$

The rate of the spontaneous transitions does not depend on $n_i$,

$$\Gamma_{sp}^{i} = \gamma_{\sigma}^{-1}|\Omega_{i}|^{2}(\bar{D}+1) \sim \eta_{i} G(\bar{D}+1). \tag{21}$$

Since $-1 \leq \bar{D} \leq 1$ the ratio $\bar{D}/(\bar{D}+1)$ is less than $\bar{D}$ for both positive and negative values of $\bar{D}$. As a consequence, the ratio $\Gamma_{st}^{i}/\Gamma_{sp}^{i}$, which is mainly determined by the product of the number of photons in the mode and the population inversion of the active medium, is bounded above:

$$\frac{\Gamma_{st}^{i}}{\Gamma_{sp}^{i}} = \frac{2n_{i}\bar{D}}{\bar{D}+1} \leq 2n_{i}\bar{D}. \tag{22}$$

As it follows from Eqs. (20) and (21), when the gain approaches zero, both rates $\Gamma_{st}^{i}$ and $\Gamma_{sp}^{i}$, as well as the number of photons in modes, tends to zero. Due to the latter, the ratio $\Gamma_{st}^{i}/\Gamma_{sp}^{i}$ also tends to zero [see Eq. (22)].

When the gain coefficient $G$ increases, the population inversion $\bar{D}$ decreases [see Eq. (15)]. Moreover, as can be seen from Eq. (15), $\bar{D}$ becomes zero at $G_{0} = N\gamma_{d}\bar{D}_{0}/\sum_{i}\alpha_{i}\eta_{i}$. As a consequence, at this gain, the ratio $\Gamma_{st}^{i}/\Gamma_{sp}^{i}$ becomes zero as well (see Appendix C). Thus, there should be two gain values at which the ratio $\Gamma_{st}^{i}/\Gamma_{sp}^{i}$ is zero. Since the ratio $\Gamma_{st}^{i}/\Gamma_{sp}^{i}$ is positive, it should achieve a maximum at some value of gain $G = G_{SE}$: $G_{th} < G_{SE} < G_{0}$. In the stationary regime, for the mode with the lowest generation threshold, the value of $G_{SE}$ can be obtained analytically (see Appendix D):

$$G_{SE} = \frac{1}{2}(\bar{D}_{0}+4)G_{th}. \tag{23}$$

For a large pump power, $\bar{D}_{0} \approx 1$, and $G_{SE} \approx 5G_{th}/2$. Our numerical solution of system (7)-(10) is in excellent agreement with the analytical evaluation Eq. (23). Indeed, as is shown in Fig. 5, the dependence of $\Gamma_{st}^{0}/\Gamma_{sp}^{0}$ on the gain reaches maximum at $G_{SE}$.

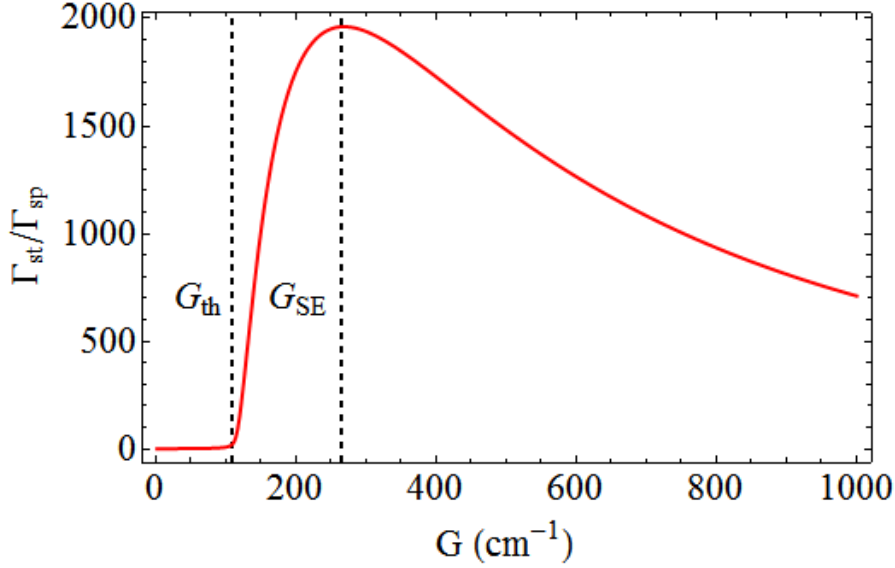

Fig. 5. The ratio of rates of stimulated and spontaneous transitions for the mode with the lowest generation threshold, $\Gamma_{st}^0/\Gamma_{sp}^0$, as a function of the gain coefficient of the active medium. The vertical dotted lines show the threshold gain coefficient, $G_{th}$, and the gain coefficient, at which the ratio $\Gamma_{st}/\Gamma_{sp}$ has a maximum.

## 6. RESULTS AND DISCUSSION

### A. The generation linewidth

When the amplification is close to the optimum, the number of photons in a mode is large, see Fig. 4. Then, the generation linewidth is determined by the Schawlow–Townes formula:[4, 57, 70]

$$\Delta \omega_i = 2\pi^2 \gamma_i \Gamma_{sp}^i(G)/\Gamma_{st}^i(G). \qquad (24)$$

The dependence of the linewidth on gain, $G$, is shown in Fig. 6. As follows from Eq.(24), the minimum of the generation linewidth coincides with the maximum of $\Gamma_{st}^i/\Gamma_{sp}^i$. Thus, for the optimal value of the gain, the linewidth is minimal. Even though near the generation threshold, $G_{th}$, the number of photons is small and the Schawlow–Townes formula is not applicable, our statement is still correct because $G_{SE} \gg G_{th}$.

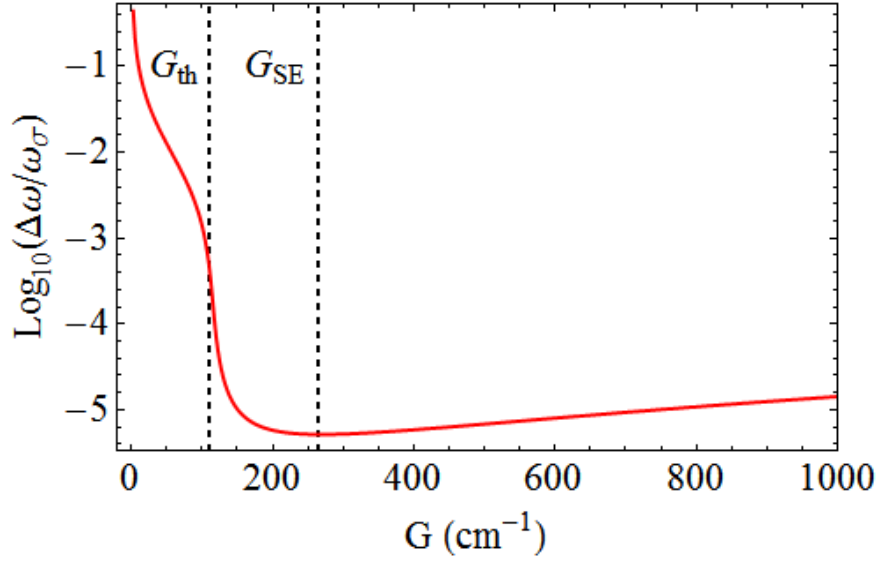

Fig. 6. The dependence of the linewidth of the lasing mode on gain.

## B. The maximum value of the photon number in the mode with the lowest generation threshold

Above its generation threshold, for the mode with the lowest threshold, the value of $\bar{D}$ is approximately the same as the threshold value $\bar{D}_{0i}^{th} = \gamma_i / \alpha_i \eta_i G > 0$ for the population inversion. We make an estimate

$$\frac{\bar{D}+1}{2\bar{D}} \approx \frac{\gamma_i + \alpha_i \eta_i G}{2\gamma_i}. \tag{25}$$

By using Eq. (21) we can express the number of photons in the $i$-th mode via the ratio $\Gamma_{st}^i / \Gamma_{sp}^i$ and the population inversion of the active medium as

$$n_i = \frac{\Gamma_{st}^i}{\Gamma_{sp}^i} \frac{\bar{D}+1}{2\bar{D}}. \tag{26}$$

Since according to Eq. (25) the factor $(\bar{D}+1)/2\bar{D}$ has no singularities near $G_{SE}$, the number of photons has a maximum near the optimal gain $G \approx G_{SE}$ where $\Gamma_{st}^i / \Gamma_{sp}^i$ has its maximum as well.

## C. The transition from a single-mode to a multimode regime

When $G$ differs from $G_{SE}$, the ratio of the rates of induced and spontaneous transitions decreases. This leads to the transition from a single-mode to a multimode regime. In the former regime, in the mode with the lowest generation threshold, the number of photons is many times greater than in other modes. This number decreases significantly when the system transitions to the multimode regime.

Let us introduce the parameter

$$\chi_i = \frac{n_i}{\sum_j n_j} \qquad (27)$$

showing the ratio of the number of photons in the $i$-th mode to the total number of photons in all modes. For the mode with the lowest generation threshold, this parameter depends on the gain coefficient non-monotonically (Fig. 7). As shown in Fig. 7, it reaches the maximum value when $G \approx G_{SE}$.

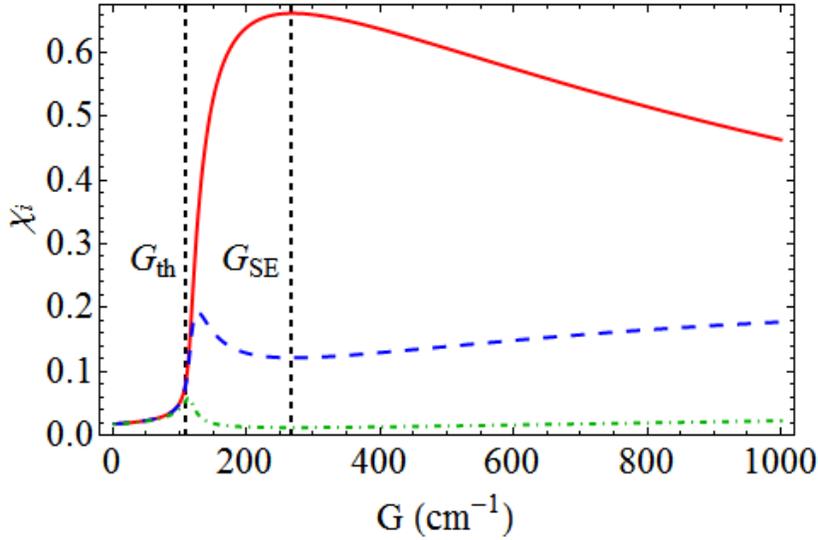

Fig. 7. The ratio of the number of photons in different modes to the total number of photons in all modes. The solid red line shows the number of photons in the mode with the lowest generation threshold ($\chi_0$), dashed blue and dash-dotted green lines show the numbers of photons in modes with the second ($\chi_1$) and the third ($\chi_2$) lowest generation thresholds, respectively.

As one can see from Fig. 7, $\chi_{i \neq 0}$ also depends on $G$ non-monotonically. Maxima of $\chi_{i=1,2}$ are near $G_{th}$. Further increase in the gain coefficient leads to a sharp increase in $\chi_0$ and a decrease in a number of photons in all other modes.

To summarize, when the gain coefficient reaches its optimum value, $G = G_{SE}$, the generation linewidth has a minimum, and the ratio of the number of photons in the mode with the lowest generation threshold to the total number of photons in all modes reaches its maximum. Thus, the system transitions into a single-mode regime. In addition, in the mode with the lowest generation threshold, the number of photons is the greatest when for $G \approx G_{SE}$.

## 7. CONCLUSION

In this paper, we have studied generation regimes of a plasmonic multimode DFB laser for various parameters of an active medium using a self-consistent model that takes into account spontaneous transitions and the multimode character of laser generation. We have shown that even when the pump power is substantially higher than threshold, the laser is multimode in a low-$Q$ plasmonic structure. In turn, the intensity and the width of the laser line depends non-monotonically on the gain coefficient of the active medium due to the nonlinear interaction of a large number of modes within the gain medium. We have demonstrated that there is an optimal value of the gain coefficient, $G_{SE} \approx 2.5 G_{th}$, at which the ratio of the rates of induced and spontaneous transitions has a maximum and the generation linewidth is at minimum.

For the optimal value of the gain coefficient, a plasmonic DFB laser supports a single-mode in which the generation occurs in the mode with the lowest generation threshold. When $G$ deviates from $G_{SE}$, the ratio of the rates of induced and spontaneous transitions decreases, and the laser works in a multimode regime. This greatly reduces the generated intensity.

Note that modes of DFB lasers are leaky waves radiating at various angles to the slab of an active medium.[24, 28, 32] Usually, the structure of a DFB laser is optimized in a way that the mode with lowest threshold radiates perpendicular to the laser plane. For $G = G_{SE}$ the ratio of the number of photons in the mode with the lowest generation threshold to the total number of photons in all modes, $\chi_0$, is the largest, $\chi_0 > \chi_i$. Thus, for $G = G_{SE}$, the plasmonic DFB laser has the narrowest radiation pattern.

The optimal value of the gain coefficient for the plasmonic DFB laser considered in our paper is $G_{SE} \approx 265\, cm^{-1}$. This value corresponds approximately to the gain coefficient of organic dye R101.[53, 54] We have used parameters of this dye in our model. Thus, one can use active media with gain coefficients of the order of a few hundred inverse centimeters in DFB lasers similar to the ones considered in our paper. These could be organic dyes or colloids of quantum dots with optical pumping or bulk semiconductors with current injection.[71]

The value of the optimal gain coefficient depends on the detuning between the eigenmode with the smallest generation threshold and the transition frequency of the gain medium. This value is inversely proportional to the ratio of the energy of the electric field of this mode in the volume of the amplifying medium to the total energy of this mode. An optimal laser can, therefore, be realized either by choosing the best active material for a given laser structure or by changing the geometry of the plasmonic structure. If the gain coefficient of the active medium is smaller than optimal, this energy ratio can be increased and vice versa.

## ACKNOWLEDGEMENT

A.A.L would like to acknowledge support from the NSF under Grant No. DMR-1312707.

# APPENDIX A: DERIVATION OF LASER EQUATIONS

In the Markovian approximation, the dynamics of two-level atoms interacting with an external electromagnetic fields may be described with a master equation in the Lindblad form[68, 69]

$$\frac{\partial \hat{\rho}}{\partial t} = -\frac{i}{\hbar}\left[\hat{H}, \hat{\rho}\right] + \hat{L}_a[\hat{\rho}] + \hat{L}_D[\hat{\rho}] + \hat{L}_\sigma[\hat{\rho}] + \hat{L}_{pump}[\hat{\rho}], \quad (A1)$$

where $\hat{\rho}$ is density matrix, $\hat{H}$ is the Jaynes–Cummings Hamiltonian[57, 68], Eq. (2). The term

$$\hat{L}_a[\hat{\rho}] = \sum_i \frac{\gamma_i}{2}\left(2\hat{a}_i\hat{\rho}\hat{a}_i^+ - \hat{a}_i^+\hat{a}_i\hat{\rho} - \hat{\rho}\hat{a}_i^+\hat{a}_i\right)$$

describes dissipation in each $i$-th mode with the dissipation rate $\gamma_i$,

$$\hat{L}_D[\hat{\rho}] = \sum_m \frac{\gamma_D}{2}\left(2\hat{\sigma}_m\hat{\rho}\hat{\sigma}_m^+ - \hat{\sigma}_m^+\hat{\sigma}_m\hat{\rho} - \hat{\rho}\hat{\sigma}_m^+\hat{\sigma}_m\right)$$

and

$$\hat{L}_\sigma[\hat{\rho}] = \sum_m \frac{\gamma_\sigma}{2}\left(\hat{D}_m\hat{\rho}\hat{D}_m - \hat{\rho}\right)$$

describe energy and phase relaxations[68, 69] with rates $\gamma_D$ and $\gamma_\sigma$, respectively. The last term in Eq. (A1),

$$\hat{L}_{pump}[\hat{\rho}] = \sum_m \frac{\gamma_{pump}}{2}\left(2\hat{\sigma}_m^+\hat{\rho}\hat{\sigma}_m - \hat{\sigma}_m\hat{\sigma}_m^+\hat{\rho} - \hat{\rho}\hat{\sigma}_m\hat{\sigma}_m^+\right)$$

describes pumping of a two-level atom at the rate $\gamma_{pump}$.[68, 69]

By using master equation (A1) and the equality $\langle \hat{A} \rangle = Tr(\hat{\rho}\hat{A})$, we can obtain a closed system of equations for expectation values of the operator of the number of photons in the $i$-th mode, $\hat{n}_i = \hat{a}_i^+\hat{a}_i$, the operator of the population inversion of the $m$-atom, $\hat{D}_m = \hat{\sigma}_m^+\hat{\sigma}_m - \hat{\sigma}_m\hat{\sigma}_m^+$, the operator of the energy flux from the $m$-th atom to the $i$-th mode,[64] $\hat{I}_{im} = -i\left(\hbar\Omega_{im}\hat{a}_i^+\hat{\sigma}_m - \hbar\Omega_{im}^*\hat{a}_i\hat{\sigma}_m^+\right)$, and the interaction operator between the $m$-th atom and the field in the $i$-th mode, $\hat{V}_{im} = \hbar\Omega_{im}\hat{a}_i^+\hat{\sigma}_m + \hbar\Omega_{im}^*\hat{a}_i\hat{\sigma}_m^+$. In the right sides of these equations, the expectation values of products of the operators appear. We have to obtain new equations for these products. This leads us to an infinite chain of equations. In order to terminate this chain, we uncouple correlators of operators of the number of photons and the population inversion, $\langle \hat{n}_i\hat{D}_m \rangle = \langle \hat{n}_i \rangle \langle \hat{D}_m \rangle$,[56] and consider that $\langle \hat{\sigma}_m^+\hat{\sigma}_l \rangle = 0$, $m \neq l$ and $\langle \hat{a}_i^+\hat{a}_j \rangle = 0$, $i \neq j$. This procedure is similar to that used in deriving the Maxwell-Bloch equations.[57] As a result, we arrive at a closed system of equations for averages $n_i = \langle \hat{n}_i \rangle$, $D_m = \langle \hat{D}_m \rangle$, $I_{im} = \langle \hat{I}_{im} \rangle$, and $V_{im} = \langle \hat{V}_{im} \rangle$ (see also Ref. 64):

$$\frac{dn_i}{dt} = -\gamma_i n_i + \sum_m I_{im}, \quad (A2)$$

$$\frac{dD_m}{dt} = (\gamma_{pump} - \gamma_D) - (\gamma_{pump} + \gamma_D)D_m - 2\sum_i I_{im}, \quad (A3)$$

$$\frac{dI_{im}}{dt} = -\left(\gamma_\sigma + \frac{\gamma_i + \gamma_{pump} + \gamma_D}{2}\right)I_{im} + (\omega_i - \omega_\sigma)V_{im} + |\Omega_{im}|^2(2n_i D_m + D_m + 1), \quad (A4)$$

$$\frac{dV_{im}}{dt} = -\left(\gamma_\sigma + \frac{\gamma_i + \gamma_{pump} + \gamma_D}{2}\right)V_{im} + (\omega_\sigma - \omega_i)I_{im}. \quad (A5)$$

Using the notations $D_m^0 = (\gamma_{pump} - \gamma_D)/(\gamma_{pump} + \gamma_D)$ and $\gamma_d = \gamma_{pump} + \gamma_D$, and taking into account that $\gamma_\sigma, \gamma_i \gg \gamma_{pump} + \gamma_D$ we obtain a system of dynamic equations governing the multimode regime of the laser

$$\frac{dn_i}{dt} = -\gamma_i n_i + \sum_m I_{im}, \quad (A6)$$

$$\frac{dD_m}{dt} = -\gamma_d (D_m - D_m^0) - 2\sum_i I_{im}, \quad (A7)$$

$$\frac{dI_{im}}{dt} = -(\gamma_\sigma + \gamma_i/2)I_{im} + (\omega_i - \omega_\sigma)V_{im} + |\Omega_{im}|^2(2n_i D_m + D_m + 1), \quad (A8)$$

$$\frac{dV_{im}}{dt} = -(\gamma_\sigma + \gamma_i/2)V_{im} + (\omega_\sigma - \omega_i)I_{im}. \quad (A9)$$

Equations (A6)-(A9) allow for describing the spontaneous emission with the accuracy of 1/$N$, where $N$ is a number of atoms.[64, 68]

## APPENDIX B: EXPRESSING CONSTANTS OF LASER EQUATIONS VIA PARAMETERS OF AN AMPLIFYING MEDIUM

The interaction constant between a field and an atom of an amplifying medium can be expressed as

$$\hbar\Omega_{im} = -\mathbf{d}_m \cdot \mathbf{E}_i(\mathbf{r}_m), \quad (B1)$$

where $\mathbf{d}_m$ is the dipole moment of the $m$-th atom at the transition frequency and $\mathbf{E}_i(\mathbf{r}_m)$ is the electric field quantum in the $i$-th mode at the position of the $m$-th atom. To normalize the electric field $\mathbf{E}_i(\mathbf{r}_m)$, we equate the energy of one quantum to the energy of the electric field in the resonator:

$$\hbar\omega_i = \frac{1}{8\pi} \int_V \left( \frac{\partial \operatorname{Re}(\varepsilon\omega)}{\partial \omega} \bigg|_{\omega=\omega_i} |\mathbf{E}_i(\mathbf{r})|^2 + |\mathbf{H}_i(\mathbf{r})|^2 \right) d^3\mathbf{r}, \tag{B2}$$

where $\varepsilon$ is the dielectric permittivity of the medium which is considered as non-magnetic. Equation (B2) gives $\mathbf{E}_i(\mathbf{r}_m)$ when the field distribution in a resonant mode is known.

If the field distribution is sufficiently uniform, the electric field in the position of each atom can be considered as approximately equal to the average field in the amplifying medium. In this case, in Eqs. (3)-(6) of the paper, one can perform the summation over the number of atoms to obtain system (7)-(10).

To calculate the coupling constant $|\Omega_i|^2$ between the field and an amplifying medium, we introduce the parameter $\eta_i$, which is the ratio of the energy of the electric field of the $i$-th eigenmode in the volume of the amplifying medium and the total energy of this mode defined by Eq. (B2):

$$\eta_i = \frac{1}{8\pi\hbar\omega_i} \int_{V_G} \frac{\partial \operatorname{Re}(\varepsilon(\mathbf{r})\omega)}{\partial \omega} \bigg|_{\omega=\omega_i} |\mathbf{E}_i|^2 d^3\mathbf{r}, \tag{B3}$$

where $V_G$ is the volume of the gain medium. Again, assuming that $|\mathbf{E}_i|^2$ is approximately constant we obtain the parameter $\eta_i$:

$$\eta_i = \frac{1}{8\pi\hbar\omega_i} V_G \frac{\partial \operatorname{Re}(\varepsilon\omega)}{\partial \omega} \bigg|_{\omega=\omega_i} |\mathbf{E}_i|^2. \tag{B4}$$

Equation (B4) expresses the average amplitude of the electric field in the gain medium via $\eta_i$:

$$|\mathbf{E}_i| = \sqrt{\frac{8\pi\eta_i\hbar\omega_i}{V_G \left[ \partial \operatorname{Re}(\varepsilon\omega)/\partial\omega \right]_{\omega=\omega_i}}}. \tag{B5}$$

Now, using Eqs. (B1) and (B5) we can find $|\Omega_{im}|^2$:

$$|\Omega_{im}|^2 = \frac{8\pi\eta_i\omega_i|\mathbf{d}|^2}{\hbar V_G \left[ \partial \operatorname{Re}(\varepsilon\omega)/\partial\omega \right]_{\omega=\omega_i}}. \tag{B6}$$

Equation (B6) allows one to obtain the coupling constant between the $i$-th resonator mode and the amplifying medium, $|\Omega_i|^2 = N|\Omega_{im}|^2$:

$$\left|\Omega_{i}\right|^{2}=\left|\Omega_{im}\right|^{2}=\frac{8\pi\eta_{i}\omega_{i}\left|\mathbf{d}\right|^{2}n_{G}}{\hbar\left[\partial\operatorname{Re}(\varepsilon\omega)/\partial\omega\right]_{\omega=\omega_{i}}}, \tag{B7}$$

where $n_G$ is the concentration of atoms of the amplifying medium. The dielectric permittivity of the amplifying medium is given by the expression:[55]

$$\varepsilon_{G}(\omega)=\varepsilon_{0}-\frac{8\pi\omega_{\sigma}n_{G}\left|\mathbf{d}\right|^{2}}{\hbar\left(\omega_{\sigma}^{2}-\omega^{2}-2i\omega\gamma_{\sigma}\right)}, \tag{B8}$$

where $\omega_\sigma$ and $\gamma_\sigma$ are the transition frequency and the rate of the transverse relaxation of a two-level atom of the amplifying medium, respectively. At the transition frequency, $\varepsilon_G(\omega)$ is

$$\varepsilon_{G}(\omega_{\sigma})=\varepsilon_{0}-i\frac{4\pi n_{G}\left|\mathbf{d}\right|^{2}}{\hbar\gamma_{\sigma}}. \tag{B9}$$

Equations (B7) and (B9) give the expression for the interaction constant $\left|\Omega_i\right|^2$:

$$\left|\Omega_{i}\right|^{2}=\left|\operatorname{Im}\varepsilon_{G}(\omega_{\sigma})\right|\frac{2\eta_{i}\gamma_{\sigma}\omega_{i}}{\left[\partial\operatorname{Re}(\varepsilon\omega)/\partial\omega\right]_{\omega=\omega_{i}}}. \tag{B10}$$

Now, using the expression for the gain coefficient[5-7, 72, 73]

$$G=-\frac{\omega}{c}\frac{\operatorname{Im}\varepsilon_{G}}{\sqrt{\operatorname{Re}\varepsilon_{G}}}, \tag{B11}$$

we can finally obtain:

$$\left|\Omega_{i}\right|^{2}=\eta_{i}G\frac{2c\gamma_{\sigma}\sqrt{\operatorname{Re}\varepsilon_{G}}}{\left[\partial\operatorname{Re}(\varepsilon\omega)/\partial\omega\right]_{\omega=\omega_{i}}}\approx\eta_{i}G\frac{2c\gamma_{\sigma}}{\sqrt{\operatorname{Re}\varepsilon_{G}}}. \tag{B12}$$

Using Eqs. (B9) and (B11) the gain $G$ may be rewritten as

$$G=n_{G}\frac{\omega}{c}\frac{4\pi\left|\mathbf{d}\right|^{2}}{\hbar\gamma_{\sigma}\sqrt{\operatorname{Re}\varepsilon_{G}}}=n_{G}\sigma_{G}, \tag{B13}$$

where $\sigma_G$ is a cross section of atom of the gain medium with the electromagnetic wave with frequency $\omega$.[5-7, 74]

# APPENDIX C: THE DEPENDENCE OF THE RATE OF INDUCED TRANSITIONS ON THE GAIN COEFFICIENT

The ratio of the rates of induced and spontaneous transitions is given by Eq. (22):

$$\frac{\Gamma_{st}}{\Gamma_{sp}} = \frac{2n_i \bar{D}}{\bar{D}+1} \leq 2n_i \bar{D}. \tag{C1}$$

Using Eqs. (15) and (18) we can estimate the right hand side of Eq. (22):

$$2n_i \bar{D} = 2n_i \frac{\gamma_d \bar{D}_0 - \frac{1}{N}\sum_j \alpha_j \eta_j G}{\gamma_d + \frac{1}{N}\sum_j \alpha_j \eta_j G(2n_j+1)}$$

$$\leq \frac{\gamma_d N \bar{D}_0}{G} \frac{n_i}{\sum_j \alpha_j \eta_j n_j} \leq \frac{\gamma_d N \bar{D}_0}{G} \frac{n_i}{\alpha_i \eta_i n_i} = \frac{\gamma_d N \bar{D}_0}{G} \frac{1}{\alpha_i \eta_i}. \tag{C2}$$

From inequalities (C1) and (C2) it follows that

$$\frac{\Gamma_{st}}{\Gamma_{sp}} \leq \frac{\gamma_d N \bar{D}_0}{G} \frac{1}{\alpha_i \eta_i} = \frac{N}{G}\left( \frac{c\gamma_d \bar{D}_0}{\eta_i} \frac{2\pi[\partial \operatorname{Re}(\varepsilon\omega)/\partial\omega]_{\omega=\omega_i}}{\operatorname{Re}\sqrt{\varepsilon_G}} \frac{(\gamma_\sigma^2 + (\omega_\sigma - \omega_i)^2)}{\gamma_\sigma^2} \right). \tag{C3}$$

From Eq. (C3) one can see that when $G \to \infty$, $\Gamma_{st}/\Gamma_{sp}$ approaches zero. In addition, an increase in the overlap integral of the mode with the gain medium $\eta_i$ and a decrease in the detuning between the transition frequency of the gain medium $\omega_\sigma$ and the eigenmode of the resonator $\omega_i$ also result in a decrease of the ratio $\Gamma_{st}/\Gamma_{sp}$. On the other hand, this ratio is not affected by a change in the concentration $n_G$ because an increase in the concentration causes a simultaneous increase in the number of active atoms $N$.

# APPENDIX D: THE OPTIMUM VALUE OF THE GAIN COEFFICIENT

First, we show that the maximum of the ratio $\Gamma_{st}^i / \Gamma_{sp}^i$ is reached for the optimum value of the gain coefficient $G_{SE}$ that corresponds to the minimum of the denominator in Eq. (14). This ratio can be expressed through variables $\phi_i = \alpha_i \eta_i G \bar{D}/\gamma_i$ as [see Eqs. (14) and (22)]

$$\frac{\Gamma_{st}^i}{\Gamma_{sp}^i} = \frac{2n_i \bar{D}}{\bar{D}+1} = \frac{2\bar{D}}{\bar{D}+1}\left(\frac{1}{2}\frac{\alpha_i \eta_i G}{\gamma_i}\frac{\bar{D}+1}{1-\alpha_i \eta_i G \bar{D}/\gamma_i}\right) = \frac{\alpha_i \eta_i G \bar{D}}{\gamma_i}\frac{1}{1-\alpha_i \eta_i G \bar{D}/\gamma_i} = \frac{\phi_i}{1-\phi_i}, \tag{D1}$$

The product $G\bar{D}(G)$ is limited due to nonlinearity of the system. Also, $G\bar{D}(G)$ cannot be greater than $G_{SE}\bar{D}(G_{SE})$ and therefore, $\phi_i = \alpha_i\eta_i G\bar{D}/\gamma_i < 1$. Then, $\Gamma^i_{st}/\Gamma^i_{sp}$ reach its maximum when $\phi_i = \alpha_i\eta_i G\bar{D}/\gamma_i$ are maximal. This happens for $G = G_{SE}$. As both gain, $G$, and the population inversion, $\bar{D}$, have the same value for each mode, $\phi_i = \alpha_i\eta_i G\bar{D}/\gamma_i$ have maxima for the same $G$ and $\bar{D}$ in every mode. Therefore, in each mode, the ratio of the rates of induced and spontaneous transitions is maximal for the optimum value of the gain coefficient. The values of these maxima for various modes can be different.

Let us now find the value of $G_{SE}$ that corresponds to the minimum of the denominator of Eq. (14), i.e. to the maximum of $\phi_0 = \alpha_0\eta_0 G\bar{D}/\gamma_0$. From Eq. (15) we obtain

$$\alpha_0\eta_0 G\bar{D} = \frac{\alpha_0\eta_0\gamma_d G\bar{D}_0 - \alpha_0\eta_0 G \frac{1}{N}\sum_i \alpha_i\eta_i G}{\gamma_d + \frac{1}{N}\sum_i \alpha_i\eta_i G(2n_i + 1)} = \gamma_0\phi_0. \tag{D2}$$

Within the linewidth of the amplifying medium, all $\alpha_i$ are approximately the same. Therefore,

$$\sum_i \alpha_i\eta_i = \alpha_0\eta_0 K, \tag{D3}$$

where $K$ is the number of modes within the linewidth. Taking into account that the greatest number of photons is in the mode with the lowest threshold, we can assume

$$\sum_i \alpha_i\eta_i n_i \approx \alpha_0\eta_0 \sum_i n_i. \tag{D4}$$

As follows from Eqs. (7) and (8), in the stationary state

$$\sum_i \gamma_i n_i = \frac{1}{2}N\gamma_d(\bar{D}_0 - \bar{D}). \tag{D5}$$

Again, keeping in mind that the greatest number of photons is in the mode with the lowest threshold, we obtain:

$$\sum_i n_i \approx \frac{1}{2\gamma_0}N\gamma_d(\bar{D}_0 - \bar{D}). \tag{D6}$$

Using Eqs. (D2)-(D4) and (D6) we arrive at

$$\alpha_0\eta_0\gamma_d G\bar{D}_0 - \alpha_0^2\eta_0^2 G^2\frac{K}{N} = \gamma_d\gamma_0\phi_0 + \alpha_0\eta_0\gamma_d G(D_0 - D)\phi_0 + \alpha_0\eta_0 G\frac{K}{N}\gamma_0\phi_0. \tag{D7}$$

Using the parameter $\phi_0$ we can rewrite $\alpha_0 \eta_0 \gamma_d GD\phi$ as $\gamma_d \gamma_0 \phi_0^2$. As a result, we obtain a quadratic equation for $\alpha_0 G$ which coefficients depend on the parameter $\phi_0$:

$$\alpha_0^2 \eta_0^2 G^2 \frac{K}{N} + \alpha_0 \eta_0 G \left( \frac{K}{N} \gamma_0 \phi_0 - \gamma_d \bar{D}_0 (1-\phi_0) \right) + \gamma_d \gamma_0 \phi_0 (1-\phi_0) = 0. \tag{D8}$$

The minimum of the denominator in Eq. (14) is achieved when

$$\left( \frac{K}{N} \gamma_0 \phi_0 - \gamma_d \bar{D}_0 (1-\phi_0) \right)^2 - \frac{4K}{N} \gamma_d \gamma_0 \phi_0 (1-\phi_0) = 0. \tag{D9}$$

Since the number of modes in the laser is much smaller than the number of atoms in the gain medium, the term with $K^2/N^2$ can be neglected. Then, we obtain

$$\gamma_d \bar{D}_0^2 (1-\phi_0) = \frac{2K}{N} \gamma_0 (\bar{D}_0 + 2) \phi_0, \tag{D10}$$

and the solution of Eq. (D8)

$$\alpha_0 G_{SE} = \frac{\gamma_0}{\eta_0} \phi_0 \left( \frac{\bar{D}_0 + 4}{2\bar{D}_0} \right). \tag{D11}$$

From Eq. (D9), in the first approximation with respect to the parameter $K/N$, we have

$$\phi_0 = \frac{\gamma_d \bar{D}_0^2}{\gamma_d \bar{D}_0^2 + \frac{2K}{N} \gamma_0 (\bar{D}_0 + 2)} \approx 1 - \frac{2K}{N} \frac{\gamma_0}{\gamma_d} \frac{\bar{D}_0 + 2}{\bar{D}_0^2}. \tag{D12}$$

Eqs. (D11) and (D12) give the optimum value of the amplification coefficient:

$$G_{SE} = \frac{1}{2} (\bar{D}_0 + 4) G_{th}. \tag{D13}$$

Note, that in deriving Eq. (D13) we assume that the number of atoms *N* remains constant when the amplification coefficient changes.